\documentclass[aps,prl,twocolumn,showpacs,floatfix,preprintnumbers,superscriptaddress,amsmath,amssymb]{revtex4-1}

\usepackage{graphicx}
\usepackage{epsfig}
\usepackage{bm}
\usepackage{color}
\usepackage{float}
\usepackage{dcolumn}
\usepackage{multirow} 
\usepackage{hyperref}
\usepackage{dcolumn}

\newcommand{\beq}{\begin{equation}}
\newcommand{\eeq}{\end{equation}}
\newcommand{\beqa}{\begin{eqnarray}}
\newcommand{\eeqa}{\end{eqnarray}}

\newcommand{\NNLOsat}{NNLO$_{\rm sat}$}
\newcommand{\NNEM}{N$^3$LO$_{\rm EM}$}

\begin{document}

\title{Accurate nuclear radii and binding energies from a chiral interaction}

\author{A.~Ekstr\"om} \affiliation{Department of Physics and
  Astronomy, University of Tennessee, Knoxville, TN 37996, USA}
\affiliation{Physics Division, Oak Ridge National Laboratory, Oak
  Ridge, TN 37831, USA}

\author{G.~R.~Jansen}
\affiliation{Physics Division, Oak Ridge National Laboratory, Oak
  Ridge, TN 37831, USA} \affiliation{Department of Physics and
  Astronomy, University of Tennessee, Knoxville, TN 37996, USA}

\author{K.~A.~Wendt} \affiliation{Department
  of Physics and Astronomy, University of Tennessee, Knoxville, TN
  37996, USA} \affiliation{Physics Division, Oak Ridge National
  Laboratory, Oak Ridge, TN 37831, USA} 

\author{G.~Hagen} \affiliation{Physics Division, Oak Ridge National
  Laboratory, Oak Ridge, TN 37831, USA} \affiliation{Department of
  Physics and Astronomy, University of Tennessee, Knoxville, TN 37996,
  USA}

\author{T.~Papenbrock} \affiliation{Department
  of Physics and Astronomy, University of Tennessee, Knoxville, TN
  37996, USA} \affiliation{Physics Division, Oak Ridge National
  Laboratory, Oak Ridge, TN 37831, USA} 

\author{B.~D.~Carlsson} \affiliation{Department of Fundamental Physics,
  Chalmers University of Technology, SE-412 96 G\"oteborg, Sweden}

\author{C.~Forss\'en} \affiliation{Department of Fundamental Physics,
  Chalmers University of Technology, SE-412 96 G\"oteborg, Sweden}
\affiliation{Department of Physics and Astronomy, University of
  Tennessee, Knoxville, TN 37996, USA} \affiliation{Physics Division,
  Oak Ridge National Laboratory, Oak Ridge, TN 37831, USA}

\author{M.~Hjorth-Jensen} \affiliation{Department of Physics and
  Astronomy and NSCL/FRIB Laboratory, Michigan State University, East
  Lansing, MI 48824, USA} \affiliation{Department of Physics,
  University of Oslo, N-0316 Oslo, Norway}

\author{P.~Navr\'atil} \affiliation{TRIUMF, 4004 Wesbrook Mall, Vancouver,
British Columbia, V6T 2A3 Canada}

\author{W.~Nazarewicz} \affiliation{Department of Physics and
  Astronomy and NSCL/FRIB Laboratory, Michigan State University, East
  Lansing, MI 48824, USA} \affiliation{Physics Division, Oak Ridge
  National Laboratory, Oak Ridge, TN 37831, USA} \affiliation{Faculty
  of Physics, University of Warsaw, Pasteura 5, 02-093 Warsaw,
  Poland}

\begin{abstract} 
  With the goal of developing predictive {\it ab-initio} capability for light and medium-mass nuclei,
  two-nucleon and three-nucleon forces from chiral effective field
  theory are optimized simultaneously to low-energy nucleon-nucleon
  scattering data, as well as binding energies and radii of
  few-nucleon systems and selected isotopes of carbon and oxygen.
  Coupled-cluster calculations based on this interaction, named
  {\NNLOsat{}}, yield accurate binding energies and radii of nuclei up
  to $^{40}$Ca, and are consistent with the empirical saturation point
  of symmetric nuclear matter. In addition, the low-lying collective
  $J^\pi=3^-$ states in $^{16}$O and $^{40}$Ca are described
  accurately, while spectra for selected $p$- and $sd$-shell nuclei
  are in reasonable agreement with experiment.
\end{abstract}


\maketitle {\it Introduction} -- Interactions from chiral effective
field theory
(EFT)~\cite{bedaque2002,epelbaum2009,machleidt2011,hammer2013} and
modern applications of renormalization group
techniques~\cite{bogner2003,navratil2009,bogner2010,furnstahl2013}
have opened the door for a description of atomic nuclei consistent
with the underlying symmetries of quantum chromodynamics, the theory
of the strong interaction.  Chiral nuclear forces can be constructed
systematically from long-range pion physics augmented by contact
interactions.  Over the past decade, the renaissance of nuclear theory
based on realistic nuclear forces and powerful computational methods
has pushed the frontier of {\it ab initio} calculations from few-body
systems and light nuclei~\cite{pieper2001,navratil2009,barrett2013} to
medium-mass
nuclei~\cite{mihaila2000b,dean2004,hagen2008,barbieri2009,hagen2010b,epelbaum2010,lahde2014,soma2013,hergert2013b}.

\begin{figure}[htb]
\includegraphics[width=\columnwidth]{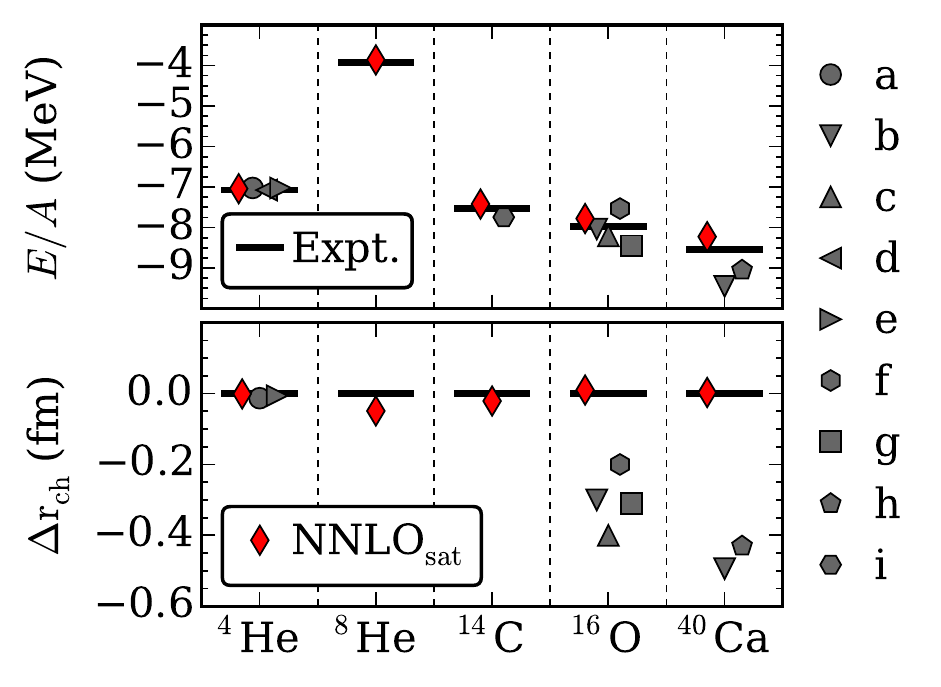}
\caption{(Color online) Ground-state energy (negative of binding
  energy) per nucleon (top), and residuals (differences between
  computed and experimental values) of charge radii (bottom) for
  selected nuclei computed with chiral interactions. In most cases,
  theory predicts too small radii and too large binding energies.
  References: $a$~\cite{navratil2007b,jurgenson2011},
  $b$~\cite{binder2013b}, $c$~\cite{epelbaum2014},
  $d$~\cite{epelbaum2012}, $e$~\cite{maris2014}, $f$~\cite{wloch2005},
  $g$~\cite{hagen2012a}, $h$~\cite{bacca2014},
  $i$~\cite{maris2011}. The red diamonds are \NNLOsat{} results
  obtained in this work.}
\label{tab_saturation}
\end{figure}

One of the main challenges in {\it ab-initio} calculations is the
accurate \footnote{In this Rapid Communication accuracy refers to an
  agreement with data at the precision one would expect from the model
  and method.}  reproduction of binding energies and radii of finite
nuclei simultaneously with the empirical nuclear matter saturation
point (binding energy per nucleon $E/A\approx 16$\,MeV at Fermi
momentum $k_F \approx 1.33$\,fm$^{-1}$) and incompressibility ($ 250 <
K_0 < 315$\,MeV \cite{stone2014}). For instance, lattice EFT
calculations at next-to-next-to leading order (NNLO) employ a
phenomenological four-nucleon contact force (a correction beyond NNLO)
to counter the overbinding in nuclei heavier than
$^{12}$C~\cite{lahde2014}, while the radii of $^{12}$C and $^{16}$O
are still too small~\cite{epelbaum2012,epelbaum2014}.  {\it Ab initio}
calculations overbind medium-mass and heavy nuclei by about 1~MeV per
nucleon, underestimate charge radii~\cite{binder2013b}, and yield too
large separation energies~\cite{hergert2014}. The status of
chiral-force predictions for binding energies and charge radii in
finite nuclei is summarized in Fig.~\ref{tab_saturation}, with dark
grey symbols representing the predictions of various state-of-the-art
calculations.  This is a serious shortcoming of current chiral
Hamiltonians as it prevents theory from making accurate predictions
when extrapolating to higher masses.  The problem with the
reproduction of nuclear matter saturation properties has been
discussed extensively in the
literature~\cite{muther2000,hh2000,dewulf2003,dickhoff2004,sammarruca2010,vandalen2010,krewald2010,hebeler2011,tsang2012,baardsen2013,carbone2013,kohno2013,hebeler2013b,shirokov2014},
and various solutions have been proposed, ranging from short-range
correlations and Pauli blocking effects to the inclusion of many-body
forces.

We start from the optimization of the chiral interaction at
NNLO. Traditionally, one takes the pion-nucleon
coupling constants $c_i$'s either from pion-nucleon scattering
\cite{epelbaum2000,buttiker2000} or from peripheral partial waves in
the nucleon-nucleon ($NN$) sector \cite{entem2003,ekstrom2013}, while
the remaining coupling constants (denoted as low-energy constants
(LECs)) are adjusted in the $NN$ sector. The corresponding $\chi^2$
optimizations consider scattering data up to laboratory energies of
$T_{\rm Lab}\approx 350$~MeV. In a subsequent step, the remaining LECs
of the leading three-nucleon ($NNN$)
forces~\cite{vankolck1994,epelbaum2002,navratil2007} are adjusted to
data on $A\le 4$ systems~\cite{navratil2007b,epelbaum2009b,gazit2009}.
For details, we refer the reader to
Refs.~\cite{epelbaum2009,machleidt2011,hammer2013}. Hitherto such a
strategy has not produced interactions that simultaneously describe
bulk properties of both nuclei and nuclear matter ~\cite{hagen2013b}.

Our optimization strategy is based on a different approach. Most
importantly, we optimize $NNN$ forces together with $NN$ forces. This
is consistent with the idea of EFT that improvements are made {\it
  order by order} and not nucleon by nucleon.  The simultaneous
optimization of $NN$ and $NNN$ forces is important because the
long-range contributions of the $NNN$ force contain LECs from
pion-nucleon vertices that also enter the $NN$ force. Moreover, in
addition to low-energy $NN$ data and the binding energies and charge
radii of $^{3}$H, $^{3,4}$He, our set of fit-observables also contains
data on heavier nuclei; namely, binding energies and radii of carbon
and oxygen isotopes.  This is a major departure from the traditional
approach that seeks to adjust LECs to data on few-body systems with
$A=2,3,4$ and then attempts to extrapolate to nuclei with $A\gg 1$ and
to infinite nuclear matter. The following arguments motivate the
strategy of including heavier nuclei into the optimization: First, no
reliable experimental data constrain the isospin $T=3/2$ components of
the $NNN$ force in nuclei with mass numbers $A=3,4$ (see
Refs.~\cite{Lauzaukas,viviani2011} for more discussion and
prospects). Second, since our goal is to describe nuclear properties
at low energies, LECs are adjusted to low-energy observables (as
opposed to the traditional adjustment of two-nucleon forces to $NN$
scattering data at higher energies). Third, the impact of many-body
effects entering at higher orders (e.g., higher-rank forces) might be
reduced if heavier systems, in which those effects are stronger, are
included in the optimization.

Besides these theoretical arguments, there is also one practical
reason for a paradigm shift: predictive power and large extrapolations
do not go together. In traditional approaches, where interactions are
optimized for $A=2,3,4$, small uncertainties in few-body systems
(e.g., by forcing a rather precise reproduction of the $A=2,3,4$
sectors at a rather low order in the chiral power counting) get
magnified tremendously in heavy nuclei, see for example
Ref.~\cite{binder2013b}.  Consequently, when aiming at reliable
predictions for heavy nuclei, it is advisable to use a model that
performs well for light- and medium-mass systems. In our approach,
light nuclei are reached by interpolation while medium-mass nuclei by
a modest extrapolation. In this context, it is worth noting that the
most accurate calculations for light nuclei with $A\le
12$~\cite{wiringa2002} employ $NNN$ forces adjusted to 17 states in
nuclei with $A\le 8$~\cite{pieper2001b}.  Finally, we point out that
nuclear saturation can be viewed as an emergent phenomenon.  Indeed,
little in the chiral EFT of nuclear forces suggest that nuclei are
self-bound systems with a central density (or Fermi momentum) that is
practically independent of mass number. This viewpoint makes it
prudent to include the emergent momentum scale into the optimization,
which is done in our case by the inclusion of charge radii for
$^{3}$H, $^{3,4}$He, $^{14}$C, and $^{16}$O.  This is similar in
spirit to nuclear mean-field calculations~\cite{gogny1970} and nuclear
density functional theory~\cite{bender2003,kortelainen2010} where
masses and radii provide key constraints on the parameters of the
employed models.

{\it Optimization protocol and model details} -- We seek to minimize
an objective function to determine the optimal set of coupling
constants of the chiral $NN$+$NNN$ interaction at NNLO.  Our dataset
of fit-observables includes the binding energies and charge radii of
$^{3}$H, $^{3,4}$He, $^{14}$C, and $^{16}$O, as well as binding
energies of $^{22,24,25}$O as summarized in
Table~\ref{tab_nnlo3NF}. To obtain charge radii $r_{\rm ch}$ from
computed poin-proton radii $r_{\rm pp}$ we use the standard
expression~\cite{friar75}: $ \langle r_{\rm ch}^2 \rangle = \langle
r_{\rm pp}^2 \rangle + \langle R_{\rm p}^2 \rangle +
\frac{N}{Z}\langle R_{\rm n}^2 \rangle +
\frac{3\hbar^2}{4m_{p}^2c^2}$, where $ \frac{3\hbar^2}{4m_{p}^2c^2} =
0.033$ fm$^2$ (Darwin-Foldy correction), $R_{\rm n}^2 = -0.1149(27)$
fm$^2$~\cite{angeli2013} and $R_{\rm p} = 0.8775(51)$
fm~\cite{mohr2010}. In this work we ignore the spin-orbit contribution
to charge radii~\cite{ong2010}. From the $NN$ sector, the objective
function includes proton-proton and neutron-proton scattering
observables from the SM99 database~\cite{SM99} up to 35~MeV scattering
energy in the laboratory system as well as effective range parameters,
and deuteron properties (see Table~\ref{tab_nnlo_observables}). The
maximum scattering energy was chosen such that an acceptable fit to
both $NN$ scattering data and many-body observables could be achieved.

\begin{table}[hbt]
  \caption{\label{tab_nnlo3NF} Binding energies (in MeV) and
    charge radii (in fm) for $^{3}$H, $^{3,4}$He, $^{14}$C and
    $^{16,22,23,24,25}$O employed in the optimization of {\NNLOsat{}}.
 }
\begin{ruledtabular}
\begin{tabular}{cdddd} 
& \multicolumn{1}{c}{$E_{\rm gs}$}&\multicolumn{1}{r}{Exp.~\cite{wang2012}}&\multicolumn{1}{r}{$r_{\rm ch}$} & \multicolumn{1}{r}{Exp. \cite{angeli2013,mohr2010}}\\ \hline
        \\[-6pt]
$^{3}$H  & 8.52       & 8.482       &  1.78       & 1.7591(363)\\
$^{3}$He & 7.76       & 7.718       &  1.99       & 1.9661(30) \\
$^{4}$He & 28.43      & 28.296      &  1.70       & 1.6755(28) \\
$^{14}$C & 103.6      & 105.285     &  2.48       & 2.5025(87) \\
$^{16}$O & 124.4      & 127.619     &  2.71       & 2.6991(52) \\
$^{22}$O & 160.8      & 162.028(57) &             &           \\
$^{24}$O & 168.1      & 168.96(12)  &             &           \\
$^{25}$O & 167.4      & 168.18(10)  &             &           \\ 
\end{tabular}
\end{ruledtabular} 
\end{table}

\begin{table}[htb]
  \caption{\label{tab_nnlo_observables}
    Low-energy $NN$ data included in the optimization. 
    The scattering lengths $a$ and effective ranges $r$ are in units of fm.
    The proton-proton observables with superscript $C$ 
    include the Coulomb force. The deuteron binding
    energy ($E_D$, in MeV), structure radius ($r_D$, in fm), and quadrupole moment ($Q_D$, in fm$^2$)  are
    calculated without meson-exchange currents or relativistic
    corrections. The computed $d$-state probability of the deuteron is 3.46\%.}
     \begin{ruledtabular} 
\begin{tabular}{cdddl} 
          &\multicolumn{1}{r}{\NNLOsat{}}&\multicolumn{1}{r}{\NNEM} \cite{entem2003} &\multicolumn{1}{c}{Exp.} & Ref. \\ \hline
$a^C_{pp}$&  -7.8258  &   -7.8188  & -7.8196(26)    &\cite{Ber88} \\
$r^C_{pp}$&   2.855    &   2.795    &  2.790(14)     &\cite{Ber88}  \\
$a_{nn}$  &     -18.929 &   -18.900    & -18.9(4)     & \cite{Chen08} \\
$r_{nn}$  &       2.911 &    2.838    & 2.75(11)       & \cite{Miller90} \\
$a_{np}$  &      -23.728&   -23.732   & -23.740(20)    & \cite{machleidt2001} \\
$r_{np}$  &      2.798  &   2.725     & 2.77(5)        & \cite{machleidt2001} \\ 
\hline \\[-6pt]
$E_D$                   &  2.22457   &   2.22458   & 2.224566 & \cite{wang2012} \\
$r_D$& 1.978     &  1.975      & 1.97535(85)  & \cite{Huber98} \\
$Q_D$     & 0.270       &   0.275     & 0.2859(3)      &  \cite{machleidt2001} 
\end{tabular}
   \end{ruledtabular} 
\end{table} 

In the present optimization protocol, the NNLO chiral force is tuned
to low-energy observables.  The comparison with the high-precision
chiral $NN$ interaction {\NNEM} \cite{entem2003} and experimental data
presented in Table~\ref{tab_nnlo_observables} demonstrates the quality
of {\NNLOsat{}} at low energies.

The results for $^{3}$H and $^{3,4}$He (and $^6$Li) were computed with
the no-core shell model (NCSM) \cite{navratil2009,barrett2013}
accompanied by infrared extrapolations~\cite{more2013}. The $NNN$
force of \NNLOsat{}\ yields about 2~MeV of binding energy for
$^4$He. Heavier nuclei are computed with the coupled-cluster method
(see Ref.~\cite{hagen2013c} and the discussion below).

A total of 16 LECs determine the strengths of the $NN$ contact
potential, the $\pi N$ potential in the $NN$+$NNN$ sector, and the
$NNN$ contacts. The LECs are constrained simultaneously by the
optimization algorithm POUNDerS~\cite{kortelainen2010}. We employ
standard nonlocal regulators in the construction of the potential, see
e.g., Refs~\cite{epelbaum2002,entem2003} for details. This type of
regulator improves the convergence of nuclear matter
calculations~\cite{hagen2013b}. In detail, the regulator functions
consist of exponentiated Jacobi momenta $p$ divided by a cutoff value
$\Lambda$, i.e., $\sim \exp[(p/\Lambda)^{2n}]$. For the present work,
we set $n=3$ and $\Lambda=450$ MeV. Furthermore, the subleading
two-pion exchange in the $NN$ interaction is regularized using
spectral function regularization with a cutoff $\Lambda_{\rm
  SFR}=700$\,MeV. The details of this procedure can be found in
Refs.~\cite{epelbaum04,epelbaum06}.

The objective function is numerically expensive, requiring us to adopt
some approximations when computing nuclei with $A>4$.  In the
optimization, we employed a model space of 9 oscillator shells for the
$NN$ interaction, the energy cutoff $E_{\rm 3max}=8\hbar\Omega$ for
the $NNN$ forces, and the coupled-cluster method in its singles and
doubles approximation (CCSD). We use nucleus-dependent estimates for
larger model spaces and triples-cluster corrections based on
Ref.~\cite{hagen2010b}.  During the optimization, we verified that
these estimates were accurate by performing converged calculations.
In our final computation of the objective function and for the results
presented in this paper, we employ much larger model
spaces and coupled-cluster methods with higher precision.

The coupled-cluster calculations are based on the intrinsic
Hamiltonian $H= T-T_{\rm cm} +V_{NN} +V_{NNN}$ to minimize spurious
center-of-mass effects~\cite{hagen2009a,hagen2010b,jansen2012}. For
the binding energies presented in this paper we employ
the $\Lambda$-CCSD(T) approximation
\cite{taube2008,hagen2010b,binder2013} in a model space consisting of
15 oscillator shells with $\hbar\Omega=22$~MeV. The $NNN$ forces are
limited to an energy cutoff $E_{\rm 3max}=16\hbar\Omega$, and
truncated at the normal-ordered two-body level in the Hartree-Fock
basis~\cite{hagen2007a,roth2012}.  We also include the leading-order
residual $NNN$ contribution to the total energy as a second-order
perturbative energy correction \cite{hagen2013b}, computed with
$E_{\rm 3max}=12\hbar\Omega$.

To compute excited states in, and around, nuclei with closed shells,
we employ equation-of-motion coupled-cluster methods
\cite{stanton1993,gour2006,jansen2011,jansen2012,shen2013,
  ekstrom2014}; these are accurate for excited states that are
generalized particle-hole excitations of low rank. For instance,
$^{14}$N is computed with the charge-symmetry breaking
equation-of-motion method from the closed sub-shell nucleus $^{14}$C,
see ~\textcite{ekstrom2014}. Similar comments apply to $^{22,24}$F. The
intrinsic charge radii are computed from the two-body density matrix
(2BDM) in the CCSD approximation \cite{shavittbartlett2009}. Benchmark
calculations of the $^4$He charge radius shows that the 2BDM result is
1\% larger than the NCSM result. Intrinsic charge densities are
computed using the one-body density matrix and correcting for the
Gaussian center-of-mass wave-function
\cite{kanungo2011,hagen2013c}. In the case of $^{16}$O, this approach
has been validated against 2BDM to four significant digits.

\begin{figure}[htb]
\includegraphics[width=\columnwidth]{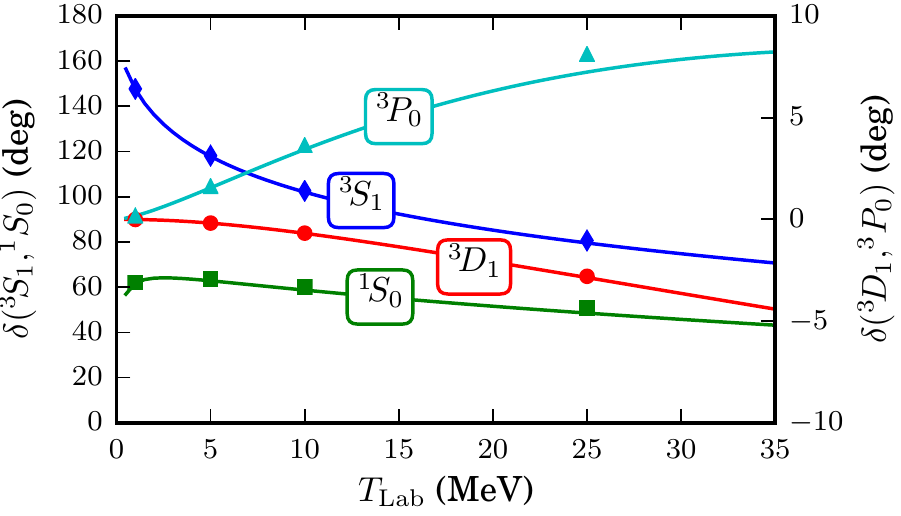}
\includegraphics[width=\columnwidth]{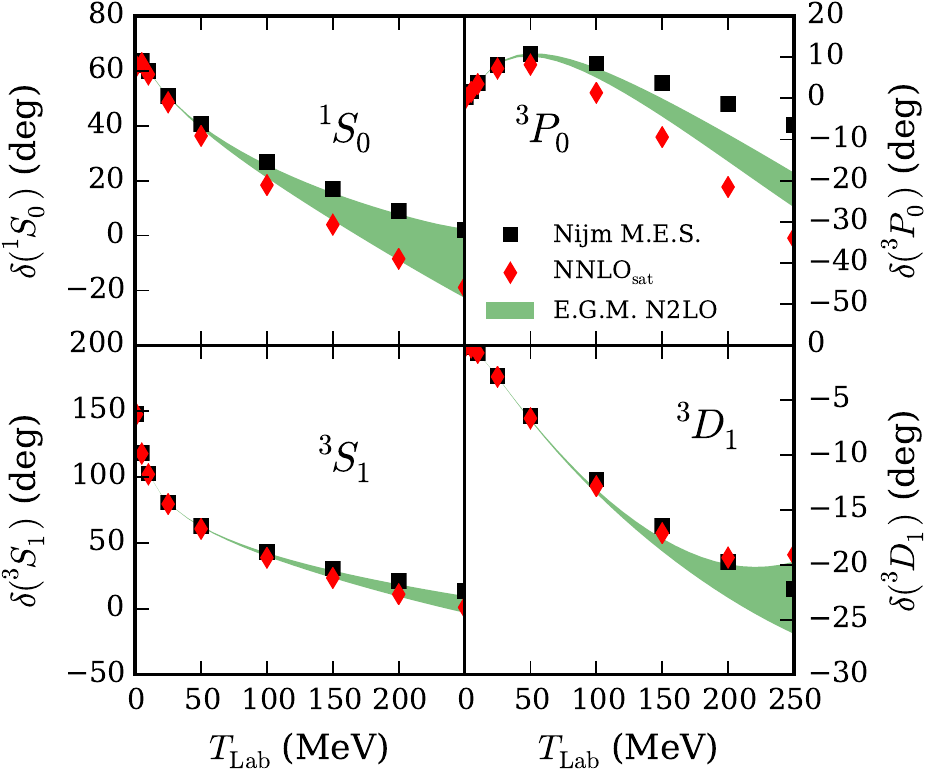}
\caption{(Color online) Selected neutron-proton scattering
  phase-shifts as a function of the laboratory scattering energy
  $T_{\rm Lab}$. (Top) {\NNLOsat{}} prediction (solid lines) compared to the
  Nijmegen phase shift analysis~\cite{stoks1993} (symbols) at low energies
  $T_{\rm Lab}<35$\,MeV. Note the
  two vertical scales. (Bottom) Neutron-proton scattering phase shifts
  from {\NNLOsat{}} (red diamonds) compared to the Nijmegen phase
  shift analysis (black squares) and the NNLO potentials (green) from
  Ref.~\cite{epelbaum04}.}
\label{fig_phases}
\end{figure}

The values for the LECs that result from the optimization and define
the chiral potential {\NNLOsat{}} are listed in Table~\ref{tab_LECS}.
\begin{table}[htb]
\caption{\label{tab_LECS} The values of the LECs for the {\NNLOsat}
  interaction. The constants $c_i$, $\tilde{C}_i$, and $C_i$ are in units
  of GeV$^{-1}$, $10^4$\,GeV$^{-2}$, and $10^4$\,GeV$^{-4}$,
  respectively.}
\begin{ruledtabular}
  \begin{tabular}{cccccc} 
LEC &  Value & LEC &  Value & LEC & Value \\ \hline   \\[-6pt]
$c_1$       & -1.12152120
 & $c_3$ & -3.92500586 & $c_4$ &  3.76568716 \\
$\tilde{C}^{pp}_{{}^1S_0}$  & -0.15814938  & $\tilde{C}^{np}_{{}^1S_0}$  & -0.15982245 & $\tilde{C}^{nn}_{{}^1S_0}$  &  -0.15915027 \\ 
$C_{{}^1S_0}$ &  2.53936779  & $C_{{}^3S_1}$ &  1.00289267 & $\tilde{C}_{{}^3S_1}$ & -0.17767436 \\
$C_{{}^1P_1}$ &  0.55595877  & $C_{{}^3P_0}$ &  1.39836559 & $C_{{}^3P_1}$ &  -1.13609526 \\
$C_{{}^3S_1-{}^3D_1}$ &  0.60071605  & $C_{{}^3P_2}$  & -0.80230030  & $c_D$ &   0.81680589 \\
$c_E$ & -0.03957471 && && 
\end{tabular}
   \end{ruledtabular}
\end{table}
We note that the pion-nucleon LECs $c_1, c_3$ and $c_4$ are in
the range of the published
values~\cite{buttiker2000,entem2003,krebs2012}. Following
Ref.~\cite{entem2003}, we set the pion-decay constant
$f_{\pi}=92.4$\,MeV and the axial-vector coupling constant
$g_A=1.29$. The value for $g_A$ is greater than the experimental
estimate $g_A=1.276$ \cite{Liu2010} to account for the
Goldberger-Treiman discrepancy. We use the following neutron, proton,
and nucleon masses: $m_{n}=939.5653$\,MeV, $m_{p}=938.272$\,MeV, and
$m_{N}=938.9184$\,MeV, respectively. For the pion masses we used
$m_{\pi^\pm}=139.5702$\,MeV and $m_{\pi^0}=134.9766$\,MeV. For the $NN$
scattering data up to 35~MeV a total $\chi^2/{\rm datum}\approx 4.3$
was reached.  Representative phase shifts are shown in
Fig.~\ref{fig_phases}. The phase shifts at higher scattering energies,
demonstrates that {\NNLOsat{}} is at the limits of expectations one
can have for an interaction at this chiral order. Furthermore, the
accuracy of \NNLOsat{} in the few-body sector is similar to other
chiral interactions at order NNLO~\cite{epelbaum2002,gezerlis2014}.

\begin{figure}[htb]
\includegraphics[width=\columnwidth]{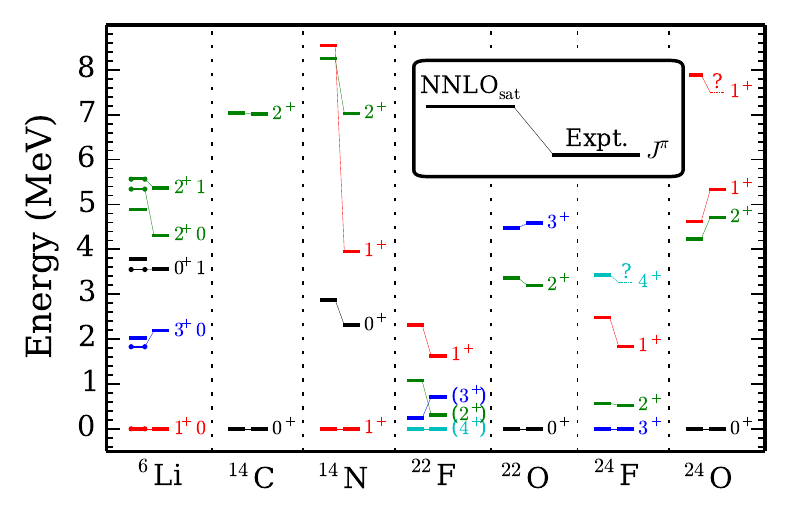}
\caption{(Color online) Energies (in MeV) of selected excited states
  for various nuclei using {\NNLOsat{}}. For $^6$Li we also include
  spectra from the NCSM (dotted lines), and isospin quantum numbers
  are also given. The NCSM results were obtained with $N_{\rm max} =
  10$ and $\hbar\Omega = 16$\,MeV. Parenthesis denote tentative spins
  assignments for experimental levels. Data are from
  Refs.~\cite{ajzenberg1991,firestone2005,firestone2007,tshoo2012}.}
\label{ex_states}
\end{figure}

{\it Predictions} -- We begin with predictions for the $\beta$-decay
half-life of $^{3}$H.  The reduced matrix element $|\langle ^{3}{\rm
  He} || E_{1}^{A} || ^{3}{\rm H} \rangle|=0.6343$ compares well to
the corresponding experimental value of $0.6848\pm
0.0011$~\cite{akulov2005,gazit2009}. Figure~\ref{tab_saturation} shows
that binding energies and charge radii of the $p$-shell nuclei $^8$He,
$^{14}$C, and $^{16}$O are in good agreement with experiment. For
$^8$He the computed binding energy and charge radius are 30.9\,MeV and
1.91\,fm, respectively, and in good agreement with the experimental
binding energy 31.5\,MeV ~\cite{wang2012} and experimental charge
radius 1.959(16)\,fm ~\cite{brodeur2012}. For $^{6,9}$Li we compute a
binding energy of 32.4(4)\,MeV and 43.9\,MeV, respectively, which
compare well with experiment (32.0\,MeV and 45.34\,MeV
~\cite{wang2012}). The charge radius of $^9$Li with \NNLOsat{} is
2.22~fm, also consistent with the measured value of 2.217(35)\,fm
\cite{sanchez2006}.  We now discuss results for excited states in
$^6$Li, $^{14}$C, $^{14}$N, and $^{16}$O, see
Fig.~\ref{ex_states}. The nucleus $^6$Li is difficult to compute
because it is bound by only 1.5\,MeV relative to the threshold for
deuteron emission. Effects of continuum is expected to lower the $2^+$
resonances significantly \cite{hupin2014}, thus we conclude that our
results are in reasonable agreement with experiment. We also compared
the spectra computed in the NCSM and agreement with the
coupled-cluster prediction is good. The binding energy computed from
two-particle attached equation-of-motion method
\cite{jansen2011,jansen2012} is 30.9\,MeV and in reasonable agreement
with the NCSM extrapolated result 32.4(4)\,MeV. Our predictions for
the excited states of $^{14}$C and $^{14}$N agree with experiment
except for the $1^+_2$ state in $^{14}$N.

The {\em ab-initio} computation of negative-parity states in $^{16}$O,
particularly the $3^-_1$ state at 6.13\,MeV
\cite{emrich1981,barbieri2002,wloch2005,epelbaum2014} has been a
long-standing theoretical challenge. We computed this state, dominated
by about 90\% of $1p$-$1h$ ($p_{1/2} \rightarrow d_{5/2}$)
excitations, at 6.34\,MeV. The energy of the $3^-_1$ state is strongly
correlated with the charge radius of $^{16}$O, with smaller charge
radii leading to higher excitation energies. For $1p$-$1h$ excited
states, the excitation energy depends on the particle-hole gap and
therefore on one-nucleon separation energies of the $A=16$ and $A=17$
systems. The charge radius depends also on the proton separation
energy $S_p$.  For $^{16}$O we find $S_p = 10.69$\,MeV and the neutron
separation energy $S_n(^{17}{\rm O}) = 4.0$~MeV, in an acceptable
agreement with the experimental values of 12.12\,MeV and 4.14\,MeV,
respectively. For $^{17}$F we find $S_p = 0.5$\,MeV, to be compared
with the experimental threshold at 0.6\,MeV. 

\begin{figure}[htb]
\includegraphics[width=\columnwidth]{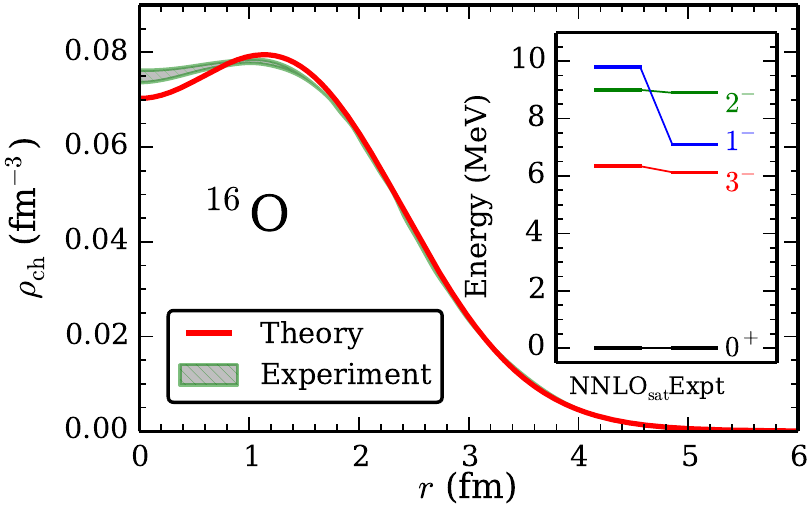}
\caption{(Color online) Charge density in $^{16}$O computed as in
  Ref.~\cite{reinhard2013} compared to the experimental charge density
  \cite{devries1987}. The inset compares computed low-lying
  negative-parity states with experiment.}
\label{ch_dens_o16}
\end{figure}

The inset of Fig.~\ref{ch_dens_o16} shows that the $2^-_1$ state in
$^{16}$O also comes out well, suggesting a $1p$-$1h$ nature. However,
the $1^-_1$ state is about 1.5~MeV too high compared to
experiment. This state is dominated by $1p$-$1h$ excitations from the
occupied $p_{1/2}$ to the unoccupied $s_{1/2}$ orbitals. In $^{17}O$
the $1/2^+$ state is computed at an excitation energy of 2.2~MeV,
which is about 1.4~MeV too high. This probably explains the
discrepancy observed for the $1^-$ state in $^{16}$O.

Figure~\ref{ch_dens_o16} shows that the experimental charge-density of
$^{16}$O is well reproduced with {\NNLOsat{}}, and our charge form
factor is, for momenta up to the second diffraction maximum, similar
in quality to what~\textcite{mihaila2000b} achieved with the Av18 +
UIX potential. For the heavier isotopes $^{22,24}$O and $^{22,24}$F
Fig.~\ref{ex_states} shows good agreement between theory and
experiment for excited states. For $^{22}$F our computed spin
assignments agree with results from shell-model
Hamiltonians~\cite{brown2006} and with recent {\it ab initio} results
\cite{ekstrom2014}. The binding energies for $^{14}$N, $^{22,24}$F are
103.7~MeV, 163~MeV and 175.1~MeV, respectively, in good agreement with
data (104.7~MeV, 167.7~MeV and 179.1~MeV). We also computed the
intrinsic charge (matter) radii of $^{22,24}$O and obtained 2.72~fm
(2.80~fm) and 2.76~fm (2.95~fm), respectively. The matter radius of
$^{22}$O agrees with the experimental result from
~\textcite{kanungo2011}. We note that the computed spectra in $^{18}$O
is too compressed compared to experiment (theory yields 0.7~MeV
compared to 1.9~MeV for the first excited $2^+$ state), possibly due
to the too high $1/2^+$ excited state in $^{17}$O. In general, the
quality of our spectra for $sd$-shell nuclei is comparable to those of
recent state-of-the-art calculations with chiral
Hamiltonians~\cite{hagen2012a,bogner2014,jansen2014,caceres2015},
while radii are much improved.

For $^{40}$Ca the computed binding energy $E=326$~MeV, charge radius
$r_{\rm ch}= 3.48$~fm, and $E(3^-_1)=3.81$~MeV all agree well with the
experimental values of 342~MeV, $3.4776(19)$~fm \cite{angeli2013}, and
3.736~MeV respectively.  We checked that our energies for the $3^-_1$
states in $^{16}$O and $^{40}$Ca are practically free from spurious
center-of-mass effects.  The results for $^{40}$Ca illustrate the
predictive power of {\NNLOsat{}} when extrapolating to medium-mass
nuclei.

\begin{figure}[htb]
\includegraphics{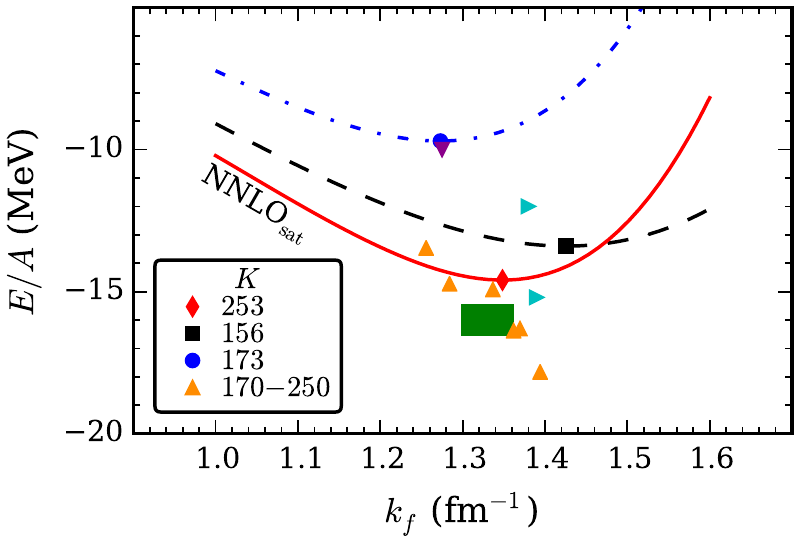}
\caption{(Color online) Equation of state for symmetric nuclear matter
  from chiral interactions. Solid red line: prediction of
  {\NNLOsat{}}.  Blue dashed-dotted and black dashed lines:
  Ref.~\cite{hagen2013b}.  Symbols (red diamond, blue circle, black
  square) mark the corresponding saturation points. Triangles:
  saturation points from other models (upward triangles
  \cite{hebeler2011}, rightward triangles \cite{coraggio2014},
  downward triangles \cite{carbone2013}). The corresponding
  incompressibilities (in MeV) are indicated by numbers. Green box:
  empirical saturation point.}
\label{snm}
\end{figure}

Finally, we present predictions for infinite nuclear matter.  The
accurate reproduction of the saturation point and incompressibility of
symmetric nuclear matter has been a challenge for {\em ab initio}
approaches, with representative results from chiral interactions shown
in Fig.~\ref{snm}. The solid line shows the equation of state for
{\NNLOsat{}}.  Its saturation point is close to the empirical point,
and its incompressibility $K=253$ lies within the accepted empirical
range \cite{stone2014}.  At saturation density, coupled-cluster with
doubles yields about 6~MeV per particle in correlation energy, while
triples corrections (and residual $NNN$ forces beyond the
normal-ordered two-body approximation) yield another 1.5~MeV.

Let us briefly discuss the saturation mechanism. Similar to $V_{{\rm
    low} k}$ potentials~\cite{bogner2003}, the $NN$ interaction of
\NNLOsat{} is soft and yields nuclei with too large binding energies
and too small radii. The $NNN$ interactions of \NNLOsat{} are
essential to arrive at physical nuclei, similarly to the role of $NNN$
forces in the saturation of nuclear matter with low-momentum
potentials~\cite{hebeler2011}. This situation is 
reminiscent of the role the three-body terms play in nuclear density
functional theory~\cite{vautherin1972}.

{\it Summary} -- We have developed a consistently optimized
interaction from chiral EFT at NNLO that can be applied to nuclei and
infinite nuclear matter. Our guideline has been the simultaneous
optimization of $NN$ and $NNN$ forces to experimental data, including
two-body and few-body data, as well as properties of selected light
nuclei such as carbon and oxygen isotopes. The optimization is based
on low-energy observables including binding energies and radii.  The
predictions made with the new interaction {\NNLOsat{}} include
accurate charge radii and binding energies.  Spectra for $^{40}$Ca and
selected isotopes of lithium, nitrogen, oxygen and fluorine isotopes
are well reproduced, as well as the energies of $3^-_1$ excitations in
$^{16}$O and $^{40}$Ca. To our knowledge, {\NNLOsat{}} is currently the
only microscopically-founded interaction that allows for a good
description of nuclei (including their masses and radii) in a wide
mass-range from few-body systems to medium-mass.

\begin{acknowledgments}
  We thank K. Hebeler and E. Epelbaum for providing the matrix
  elements of the non-local threebody interaction. This material is
  based upon work supported by the U.S. Department of Energy, Office
  of Science, Office of Nuclear Physics under Award Numbers
  DEFG02-96ER40963 (University of Tennessee), DE-SC0008499 and
  DE-SC0008511 (NUCLEI SciDAC collaboration), the Field Work Proposal
  ERKBP57 at Oak Ridge National Laboratory and the National Science
  Foundation with award number 1404159.  It was also supported by the
  Swedish Foundation for International Cooperation in Research and
  Higher Education (STINT, IG2012-5158), by the European Research
  Council (ERC-StG-240603), by the Research Council of Norway under
  contract ISP-Fysikk/216699, and by NSERC Grant No.
  401945-2011. TRIUMF receives funding via a contribution through the
  National Research Council Canada. Computer time was provided by the
  Innovative and Novel Computational Impact on Theory and Experiment
  (INCITE) program. This research used resources of the Oak Ridge
  Leadership Computing Facility located in the Oak Ridge National
  Laboratory, which is supported by the Office of Science of the
  Department of Energy under Contract No.  DE-AC05-00OR22725, and used
  computational resources of the National Center for Computational
  Sciences, the National Institute for Computational Sciences, the
  Swedish National Infrastructure for Computing (SNIC), and the Notur
  project in Norway.
\end{acknowledgments}

\bibliography{thomas}
\bibliographystyle{apsrev}

\end{document}